(Draft of October 24, 2001: To Be Replaced with Final Version)

# Asymmetric Regulation on Steroids:
# U.S. Competition Policy and Fiber to the Home

**Sharon Eisner Gillett**
Executive Director and Research Associate
MIT Program on Internet & Telecoms Convergence
sharoneg@mit.edu

**Emy Tseng**
Research Assistant
MIT Program on Internet & Telecoms Convergence
emy@itc.mit.edu

The authors gratefully acknowledge the support of the industrial sponsors of the
MIT Program on Internet & Telecoms Convergence (ITC), listed on ITC's web site at
http://itc.mit.edu.

MIT Program on Internet & Telecoms Convergence
Center for Technology, Policy and Industrial Development
Massachusetts Institute of Technology
One Amherst St, E40-234
Cambridge, MA 02139
617 253 4138
http://itc.mit.edu

# Abstract

Fiber to the Home (FTTH) describes a set of emerging technologies with the potential to affect the competitive landscape in local access. On the one hand, the high cost of deploying fiber and associated equipment all the way to the residence suggests limitations on the extent of facilities-based competition among FTTH networks. On the other hand, FTTH opens up new possibilities for service-level competition, defined as the sharing of a single network infrastructure by multiple higher-layer service providers, whether of the same or different services. The high and ever-increasing transmission capacity of fiber, and the lack of legacy allocations of that capacity in new network designs, together present a much greater degree of technical flexibility for network sharing than is found in the legacy telephone and cable networks primarily used for local access today.

Yet technology is hardly an exogenous factor that independently shapes future local access competition; the regulatory environment also plays a key role. By shaping expectations about future competitive requirements, current regulations influence network operators' deployment choices among competing FTTH technologies, as well as design choices made by vendors and standards bodies for technologies still under development.

The current regulatory approach to FTTH is far from consistent. Network operators trialing FTTH currently include Incumbent Local Exchange Carriers (ILECs), competitive access providers of various sorts (including CLECs), rural/independent telephone companies, and municipalities. This paper reviews the rules related to service-level competition that apply to each of these institutional categories, as well as to incumbent cable operators that are also likely to play a future role in FTTH. These rules vary from the detailed FCC orders on resale and unbundled network elements (UNEs) that apply to ILECs, to varying state-level restrictions on entry that apply to municipalities. In essence, the paper finds that if current regulatory trends continue, asymmetries in the regulation of service-level competition will be on steroids by the time FTTH starts being more commonly deployed. This finding is especially problematic if fiber's high deployment costs and potential for integrated services lead to highly monopolistic local access markets.

We view the high-level problem as a lack of clear regulatory commitment to service-level competition as a long-term policy goal, independent of the specific local access technology or the nature of the particular service(s) offered. Current regulatory requirements are either non-existent, or extremely detailed and technology- and service-specific (e.g. UNEs). We argue that neither of these approaches is likely to achieve the desired result for FTTH, given the current state of flux in emerging FTTH technology.

## Introduction

Fiber to the Home (FTTH) describes a set of emerging technologies whose common element is the deployment of a fiber-optic line all the way to the residential customer's premises. Although some of these technologies have been talked about for at least 10-20 years,[1] the emergence of a significant market for residential Internet access has breathed new life into the discussions. Practical interest in FTTH is resurgent, based on now-proven consumer demand for residential broadband Internet access, the continuing development and popularity of ever more bandwidth-hungry Internet applications such peer-to-peer file sharing and video distribution, and ongoing price-performance improvements in optical networking technology. In the context of these trends, FTTH is viewed by many stakeholders as the soundest basis for a more future-proof local access infrastructure.

In the second section of this paper, we explain why we see FTTH as the eventual future of residential broadband infrastructure, at least for fixed (wired) networks. We review current experiments with FTTH in the U.S. and discuss the motivations of organizations currently or potentially deploying FTTH. These organizations include incumbent telephone and cable operators, new entrants building alternative facilities, and municipalities, typically in the form of municipal utilities.

The likely emergence of FTTH networks makes it important to consider what effect they may have on future local access competition. On the one hand, the high cost of deploying fiber and associated equipment all the way to the residence suggests limitations on the extent of facilities-based competition (Reed, 1991). As with today's residential broadband Internet access, duopoly would be the best outcome, in communities in which both cable and telephone networks migrate to FTTH. Monopoly is also a likely outcome, however, because FTTH can create a significant first mover advantage. Fiber's high bandwidth capacity means that one network operator can provide multiple services (e.g. data, voice and video), leaving less room for entrants to offer novel services to attract potential customers. By this argument, the first network to migrate, or the first network to be built from scratch, precludes the ability of would-be competitors to recoup their investment in FTTH.[2]

On the other hand, FTTH opens up new possibilities for service-level competition, defined as the sharing of a single network infrastructure by multiple higher-layer service providers, whether of the same or different services. Emerging FTTH technologies present a degree of technical flexibility much greater than that of the legacy telephone and cable networks primarily used for broadband today. The high and ever-increasing transmission capacity of fiber, and the lack of legacy allocations of that capacity in new network designs, mean that FTTH capacity can potentially be shared not just among multiple users and services, but also among multiple service providers.[3]  For example, customers might buy Internet access from several different ISPs, all delivering service over the same FTTH network.  Or, they might buy Internet access and video (e.g. cable TV) services over the same network, but from different companies.  Yet another form of service-level competition might arise if the data transmission capabilities delivered over FTTH are high enough to support new modes of usage, such as switched digital video[4] and video telephony, that indirectly compete with more traditional cable TV and telephony services.

---

[1] See, for example, (Reed 1991) and the references therein.
[2] Fiber need not run all the way to the home for a network operator to offer integrated services. However, an all-fiber network offers the greatest capacity and flexibility for combining multiple services.

Yet technology is hardly an exogenous factor that independently shapes future local access competition.  The regulatory environment also plays a key role.  By shaping expectations about future competitive requirements, current regulations influence network operators' deployment choices among competing FTTH technologies.  At a more basic level, regulatory expectations also shape design choices made by vendors and standards bodies for technologies still under development.  Thus current regulations influence how likely it is that FTTH's potential to support more service-level competition will actually be realized.

The current regulatory approach to FTTH is far from consistent.  Requirements for support of service-level competition vary tremendously depending on the type of service (voice, data or video) and the identity of the network operator.  The third section of this paper reviews the details of this heterogeneity.  The paper concludes with a discussion of why this inconsistent approach is problematic, and potential policy directions that could improve this situation.

## FTTH: The Future of Residential Broadband

Despite much hype, actual deployments of FTTH in the U.S. remain at the relatively rare stage, in which TPRC authors can still list a reasonable sample of trials in a short table (see Table 1).

---

[3] See (Tseng 2001, Chapter 4) for further detail on this point.
[4] Also known as Video over IP.  See (Abe 2000) pp.334-336 for a description of Switched Digital Video.

| Type of Organization | Provider | Location(s) | Status | Technology |
|---|---|---|---|---|
| Alternative facility provider | Clearworks.net | TX (new developments) | Announced deployments. Unknown status. | Ethernet over Active Star (FTTH) + HFC (for video) |
| Alternative facility provider (CLEC) | Western Integrated Networks, LLC. (WINFirst) | Austin, TX;  San Antonio, TX; Sacramento, CA | Planning network deployments in Austin, San Antonio, Sacramento | Ethernet over Active Star (FTTH) + HFC (for video) |
| Alternative facility provider | HomeFiber | Palo Alto, CA (new developments) | Unknown | Ethernet over Active Star |
| Alternative facility provider | FiberHood | Palo Alto, CA | Running trial | Ethernet over PON |
| Alternative facility provider | Vialight | Redmond, WA (new developments) | Currently offering services | Ethernet over Active Star |
| ILEC | BellSouth | Dunwoody, GA | Running trial | ATM over PON |
| ILEC | Verizon | Not public | Running trial | Ethernet over PON |
| Rural/Independent Phone Company | Blair Telephone Co., Huntel Engineering | Blair, Nebraska (new developments) | Currently offering services | ATM over PON |
| Rural/Independent Phone Company | Rye Telephone Co. | Colorado City, CO (new developments) | Currently offering services | ATM over PON |
| Municipal Utility | City of Palo Alto Utilities | Palo Alto, CA | Building trial network | Ethernet over PON |
| Municipal Utility | Grant County PUC | Grant County, WA | Currently offering services | Ethernet over Active Star |

**Table 1: U.S. FTTH Deployments**

Several trends, however, point to FTTH becoming less rare in the foreseeable future:

1. **Existing network operators are pushing fiber ever closer to subscribers.** In order to reduce maintenance costs, increase capacity and support new services such as broadband Internet access, cable TV and telephony, cable and telephone network operators have been deploying more fiber and pushing it closer to the subscriber. For example, the Hybrid Fiber Coax (HFC) networks generally built out by cable operators in the mid-late 1990's typically brought fiber to a "neighborhood" (FTTN) of about 500-2,000 homes (Gillett, 1997). Recent overbuilders such as RCN typically bring fiber to the "curb" (FTTC), serving clusters of 20-200 homes. A logical result of this evolution is for fiber eventually to reach all the way to the customers' premises.

2. **Fiber may already be more economical than copper for greenfield sites.** Because a large portion of the costs of a new network lie in its installation,[5] in new developments the cost differential to deploy fiber all the way to the end user may be negligible. It may even be favorable when the comparison factors in life cycle (operational) costs, and/or the option that an all-fiber network grants by giving providers more flexibility to meet the uncertainties of future demand.[6] The value of this option can be realized with a variety of incremental strategies, including laying fiber to the end user premises but not making it visible to subscribers in the form of new services, thereby reducing the cost of end-user equipment needed in advance of proven demand. Such strategies are also being adopted in areas where existing networks are being largely rebuilt as part of routine maintenance.

3. **Communications infrastructure is increasingly seen by communities as essential to their economic development.** Demand for advanced telecommunications services is hardly limited to high-density urban communities, yet these are the areas where the market brings them soonest (Gillett and Lehr, 1999; Gabel and Kwan, 2000). More sparsely populated areas have come to view availability delays as a disadvantage that threatens their ability to attract and retain businesses and residents.[7] In some cases, this rationale has led communities to build their own local access network, just as they assume responsibility for water, roads and other infrastructure. When the goal is community economic development rather than profit, FTTH's advantages in providing advanced services and future-proofing are especially appealing.

These trends are evident in Table 1. About half of the deployments listed target new housing developments. Grant County, WA and Palo Alto, CA lead the trend toward municipally developed networks.[8]

Of additional interest in Table 1 are the different types of organizations involved in deploying FTTH networks. Potential FTTH network providers include the incumbent cable and telephone operators as well as new entrants such as competitive local exchange carriers (CLECs), other competitive access providers, utilities and municipalities. Motivations and strategies for deployment vary greatly among the different providers and deployment scenarios. The rest of this section reviews the trend toward FTTH within each of these categories.

---

[5] FTTH will primarily be deployed where the provider can avoid trenching—for example, in new builds, where trenches are already being dug for other utilities, and in areas where access to aerial plant is available.

[6] [*** Insert citation(s) here. ***]

[7] See, for example, the rationales for the various "Connect" projects to develop alternative broadband infrastructure in New England. These projects are linked from http://www.mtpc.org, the web site of the Massachusetts Technology Collaborative.

[8] A number of other municipalities have undertaken telecommunications infrastructure projects; these are the only two communities we are aware of that have brought fiber all the way to the home. See (Strover and Berquist, 1999).

[* Following sections to be edited for consistency *]

## ILEC

Telephone companies have several motivations to replace the existing telephone infrastructure with Fiber to the Home including lower maintenance costs and the ability to provide services such as very high speed data access and video that are not well supported by the current telephone network. Also, as the old infrastructure needs repair, telephone companies are incrementally replacing the copper wiring with fiber. However, currently advanced services are not offered even though the infrastructure may be upgraded to fiber, because of the high costs of replacing the customer equipment.

A couple of ILECs have deployed FTTH trial networks. Bell South has a trial serving 400 homes in a suburb of Atlanta, and Verizon is running a FTTH trial in an undisclosed location on the East Coast.

## Alternative facility provider

Competitive providers have varying strategies and business models for the deployment of Fiber to the Home networks. The different providers vary by the type of services they provide. Some competitive providers provide only data services. These companies are known as Competitive Access Providers.

Other companies provide a full set of services over the network including telephone, cable TV, and sometimes, advanced services such as neighborhood Intranet and video on demand. A CLEC is a provider that is licensed to provide telephone service, and is bound by the common carriage rules in the Telecommunications Act that apply to local exchange carriers outlined above.

These providers often target greenfield sites. For example, companies such as ClearWorks.net and ViaLight offer voice, video and data services to new developments, teaming up with a real estate developer to provide the network to homes in a private development. Usually the developer gives them some kind of guarantee for the number of homes that will hook up to the network. In some cases, they may get exclusive rights for a period of time. In this way they gain first mover advantage.

Companies known as overbuilders deploy networks in existing neighborhoods providing a third wire into the home. Most of these companies are deploying HFC systems even though network installation costs do not differ significantly whether they are deploying coaxial cables or fiber. However, the total cost of Fiber to the Home is still too expensive for most of these providers. FTTH will become more viable for overbuilders as equipment costs come down. One exception is Western Integrated Networks (WINFirst), which has obtained cable licenses in various cities in Texas and California announced plans is to build a combined HFC and FTTH network where coaxial cable carries the television signals and fiber carries data and voice into the home.

## Rural/independent telco

Some independent telephone companies in rural areas have adopted Fiber to the Home technology. For example, Rye Telephone Company in Colorado and Huntel Engineering offer cable TV, telephone and Internet access to their customers in new housing developments.[9]

## Municipal utility

Utilities (especially electric) are looking to broaden their business strategy into telecommunications, especially since they often have much of the infrastructure already in place. In fact, many of these organizations have already deployed fiber networks connecting their own facilities. Some lease dark fiber to other organizations and businesses. Some utilities want to build out their networks to individual homes and businesses to support applications such as automated metering. Since these applications do not take much bandwidth, they can use these networks to offer telecommunications services.

Utilities have certain advantages over other potential competitive providers. For one thing, they already have staff with the expertise in the deployment and maintenance of fiber networks. They have access to the poles and rights-of-way. They have relationships with existing customers and also have billing support systems in place. However, utilities lack experience in telecommunications, so they may partner with another company to offer telecommunications services. For example, overbuilders may partner with electrical utilities in order to obtain rights-of-way. RCN's partnership with Boston Edison is an example of this type of partnership.

In order to stimulate competition, some local governments are directly investing in telecommunications infrastructure, including local access networks. These municipalities form public/private partnerships to offer services over these facilities. These local governments are building competitive alternatives to the incumbent telephone and cable TV infrastructure.

Municipalities build telecommunications infrastructure for reasons beyond pure profit. Municipalities upgrade telecommunications infrastructure with the hope of attracting residents and businesses to their communities by offering access to advanced telecommunications services. Some municipalities, especially in rural communities, lack advanced services and feel they cannot rely on the private sector to provide them. These communities have a vested interest providing choice for their citizens, and therefore are interested in promoting service level competition over their networks.

---

[9] Huntel Engineering http://www.huntelengineering.com and Rye Telephone Company http://www.fone.net/soco/guide/colocity/rtc/home.html

Certain municipalities -- such as Grant County, Washington and Palo Alto, California— are deploying Fiber to the Home networks in their communities, extending existing publicly owned networks into the local loop themselves and partner with private companies to provide the services.

## Incumbent cable operators

Cable television providers have shown less interest than the incumbent telephone companies to deploy fiber all the way to the home. Cable television companies are taking a gradual approach to deploying fiber closer to the subscriber. Cable companies such as AT&T are migrating their HFC systems to Fiber to the Curb to allow fewer subscribers to share the network. Eventually, cable companies may migrate to all fiber networks if application demand warrants it (especially for upstream bandwidth), if incrementally upgrading the network becomes more expensive than deploying fiber all the way, or if FTTH offers significant enough savings in operational costs compared with HFC networks.

## <u>U.S. Regulatory Requirements for Service-Level Competition</u>

The current regulatory approach to service-level competition is far from consistent, varying by the type of provider and the types of services they offer (Table 2).

**Table 2: U.S. Requirements for support of service-level competition**

| | | Type of Service | | |
|---|---|---|---|---|
| | | **Voice** | **Data** | **Video** |
| **Type of Provider** | **ILEC** | UNEs, collocation and resale (TA'96 §251c) | • UNEs, collocation and resale (TA'96 §251c)<br>• Line sharing, DSL UNEs (FCC Report & Orders 3 & 4)<br>• Separate subsidiary: not (merger conditions invalidated by courts; but, watch PA) | 3 choices under TA'96 §302<br>• None ("cable"): just usual broadcast and programming rules<br>• Hybrid: "open video"<br>• VDT: "common carriage video" |
| | **Incumbent cable operator** | Allow access to rights of way, don't prohibit resale, etc. ("CLEC" rules: TA'96 §251b) | • Statutory: none<br>• Court rulings: none<br>  • ATT v. Portland: locality can't require, but FCC can<br>  • MediaOne v. Broward County: open access violates 1st Amendment<br>• Merger conditions<br>  • AOL/TW: FTC consent decree (5 years)<br>  • ATT/TCI/MediaOne: none | None (1984, 92 cable acts; ineffective "leased access") |
| | **Rural telco** | None (TA'96 §251f exemptions pre-empt §251c)[10] | None (TA'96 §251f exemptions pre-empt §251c) | None |
| | **Alternative facility provider** | Allow access to rights of way, don't prohibit resale, etc. ("CLEC" rules: TA'96 §251b) | None | None |
| | **Municipality (typically through electric utility)** | Unclear whether even allowed (differing state laws, pending court cases) | None (although may be locally required, *de jure* or *de facto*) | None (although may be locally required, *de jure* or *de facto*) |

---

[10] Cable-telephone cross-ownership restrictions are also lifted for rural telephone companies. We speculate that companies that are already allowed to provide both video and voice services might be more likely to offer integrated services over a future FTTH network.

**Voice**

[* To be filled in with explanation of Table 2 *]

**Data**

[* To be filled in with explanation of Table 2 *]

**Video**

[* To be filled in with explanation of Table 2 *]

## <u>Competition Policies for Fiber to the Home</u>

As shown in the previous section, separate regulatory regimes would apply to Fiber to the Home depending on the different providers (cable vs. telephone) and the services they offer. However, this regulatory asymmetry is increasingly untenable. As fiber gets deployed deeper into the network, the networks are becoming more similar, both in the transmission media and the services they support. Although asymmetric regulation may need to exist in order to level the playing field between various market players, the potential differences should reflect differences based on market position (incumbent vs. competitor) rather than increasingly irrelevant technological distinctions.[11]

There are issues with various specific regulatory approaches as well. The unbundling rules that apply to the telephone network are unworkable with rapidly changing technology such as FTTH. The unbundling requirements that currently apply to the telephone network are difficult to implement effectively in the context of rapidly changing technologies such as FTTH. Unbundling FTTH networks will require definition of new unbundled network elements (UNEs). However, many different FTTH technology options exist, and no dominant technology has emerged, making it difficult to define the network elements much less deciding which should be unbundled. It will also be difficult to set costs, since the network costs will vary with changing technology. Furthermore, the most detailed rules are very technology specific. It is also possible that the incumbents would architect their networks so as to circumvent these policy restrictions.

The current FCC's approach taken towards the cable broadband is reactive rather than proactive: the government is waiting to see if competition develops before imposing regulation. However, the problem is that if competition does not develop, any policy would have to be enforced within the confines of an anti-competitive market structure and technology, restricting the effectiveness of the policy. An alternative is the "laissez-faire" approach, where the government would remove the network sharing requirements from the incumbent providers for the deployment of Fiber to the Home, in order to encourage facilities-based competition. However, due to the entrenched position of the incumbent monopolists, laissez faire will not lead to increased competition.

---

[11] See (Perucci and Cimatoribus 1997) for a discussion of when asymmetric regulation should be applied.

## Conclusions

[* Insert policy recommendation and concluding discussion *].

Policy makers need to define a consistent policy direction, in advance of technical development in order to shape technical choices now, rather than grafting them on later. The policy goals regarding service level competition should be set while Fiber to the Home technology is still being developed. Once these requirements are defined, Fiber to the Home technology can be designed in such a way that increases the opportunity for service level competition. Different architectures make network sharing harder or easier (Table 3).

**Table 3: FTTH Technology Support for Service-Level Competition**

| Architecture | Description | What's Shared |
|---|---|---|
| HomeRun | Dedicated fiber per subscriber (direct connection between subscriber and meet point) | Meet point. Customer chooses how to use fiber and whom/what to connect to. |
| Active Star | Signals switched at node between user and meet point (e.g. Ethernet) | Between meet point and node. |
| Passive Star | Signal's power split at node between user and meet point (virtual bus architecture, e.g. FSAN/ATM PON) | Between meet point and user. |
| WDM PON | Evolving: Dedicated wavelength per … (service, provider, or subscriber) | Depends. Frequency unbundling, maybe. |

## Bibliography

Abe, George. 2000.  *Residential Broadband.*  2$^{nd}$ ed.  Indianapolis: Cisco Press.

*AT&T, et al.* v. *City of Portland*, U.S. Court of Appeals, Ninth Circuit, Appeal No. 99-35609. June 22, 2000.

Federal Communications Commission. 2000. "Inquiry Concerning High-Speed Access to the Internet Over Cable and Other Facilities." Notice of Inquiry.  FCC 00-355.  GN Docket No. 00-185.

Gabel, David and Florence Kwan. 2000.  "Accessibility of Broadband Communication Services by Various Segments of the American Population."  Alexandria, VA: Telecommunications Policy Research Conference.

Gillett, Sharon E. 1997. "Connecting Homes to the Internet: An Engineering Cost Model of Cable vs. ISDN." In Alain Dumort and John Dryden, eds., *The Economics of the Information Society,* Luxembourg: European Communities by arrangement with the OECD.

Gillett, Sharon E. and William H. Lehr.  1999.  "Availability of Broadband Internet Access: Empirical Evidence."  Alexandria, VA: Telecommunications Policy Research Conference.Pearah, David. 1998.  "ADSL Deployment: Law, Economics, and Strategy." Alexandria, VA: Telecommunications Policy Research Conference.

Reed, David P. 1991. *Residential Fiber Optic Networks: An Engineering and Economic Analysis*. Boston: Artech House.

Shaw, James. 1998. *Telecommunications Deregulation*. Boston: Artech House.

Sidak, Gregory (ed.), 1999. *Is the Telecommunications Act of 1996 Broken? If So, How Can We Fix It?*  AEI Press.

Strover, Sharon and Lon Berquist.  1999.  "Telecommunications Infrastructure Development: The Evolving State and City Role in the United States."  Newcastle upon Tyne: Cities in the Global Information Society: An International Perspective.

Telecommunications Act of 1996. Pub. L. No. 104-104, 110 Stat. 56 1996. Sections 103, 251, 302, 303, 652

Tseng, Emy. 2001. "Competition in Fiber to the Home: A Technology and Policy Assessment." M.S. Thesis.  Massachusetts Institute of Technology.  September, 2001.

Vogelsang, Ingo and Bridger M. Mitchell, 1997. *Telecommunications Competition: The Last Ten Miles*.  Cambridge: MIT Press.